\documentstyle[11pt,moriond,epsfig]{article}

\def\be{\begin{equation}}
\def\ee{\end{equation}}
\def\bea{\begin{eqnarray}}
\def\eea{\end{eqnarray}}

\def\lsim{\, \lower2truept\hbox{${< \atop\hbox{\raise4truept\hbox{$\sim$}}}$}\,}
\def\gsim{\, \lower2truept\hbox{${> \atop\hbox{\raise4truept\hbox{$\sim$}}}$}\,}

\begin{document}
\vspace*{4cm}
\title{EXTRAGALACTIC POINT SOURCES AND THE PLANCK SURVEYOR MISSION}

\author{ G. DE ZOTTI, L. TOFFOLATTI, G.L. GRANATO }

\address{Osservatorio Astronomico di Padova, \\
Vicolo dell'Osservatorio 5, I-35122 Padova }

\maketitle\abstracts{We review estimates of small scale fluctuations due to 
extragalactic point sources in the Planck Surveyor frequency bands. 
While our 
undestanding of the spectral and evolutionary properties of these 
sources is far from complete, conservative estimates 
allow us to confidently conclude that, in the 
frequency range 100--200 GHz, their contaminating effect 
is well below the expected anisotropy level of the cosmic microwave 
background (CMB), down to angular scales of at least 
$\simeq 10'$. Hence, an accurate subtraction of foreground fluctuations 
is not critical for the determination of the CMB power spectrum up 
to multipoles $\ell \simeq 1000$. In any case, Planck's wide frequency coverage 
will allow to carefully control foreground contributions.
On the other hand, the all sky surveys at 9 frequencies, spanning the 
range 30--900 GHz, 
will be unique  in providing complete samples comprising from several 
hundreds to many thousands of extragalactic sources, selected in an 
essentially unexplored frequency region. New classes of sources may be 
revealed in these data. The familiar ``flat''-spectrum radio sources 
should show spectral features carrying essential information on their physical 
properties. Crucial information will be provided to understand the 
nature of radio sources with strongly inverted spectra. Scenarios for 
the cosmological evolution of galaxies will be extensively tested.}

\section{Introduction} 

The multifrequency all-sky maps produced by the Planck Surveyor  
mission will comprise, in addition to anisotropies which are 
outgrowths of primordial fluctuations, and whose 
precision measurements are the main goal of the mission, astrophysical 
foregrounds, the most important of which, over the frequency range of 
interest, are those due to emissions in our own Galaxy and to 
extragalactic radio and mm/sub-mm sources.
   
We deal here with extragalactic sources, which may be a major limiting 
factor for experiments, like Planck, aimed at accurately 
determining the cosmic microwave 
Background (CMB) power spectrum $C_\ell$ up to multipoles $\ell \sim 2000$, 
corresponding to angular scales $\theta \sim 5'$.
In fact, a Poisson distribution of sources produce a white noise power spectrum 
with the same power in all multipoles~\cite{Tegmark}, 
so that their contribution to fluctuations in a unit logarithmic 
multipole interval increases with $\ell$ as $\ell(\ell +1)C_\ell \propto 
\ell^2$ (for large values of $\ell$),  
while, at least for the standard inflationary models, 
which are consistent with the available anisotropy detections,  
the function $\ell(\ell +1)C_\ell$ yielded by primordial CMB fluctuations 
is approximately constant for $\ell \lsim 100$, then oscillates and finally 
decreases quasi exponentially for $\ell \gsim 1000$  
($\theta \lsim 10'$).
Hence confusion noise due to discrete sources will 
dominate at small enough angular scales.

In \S$\,$2 we summarize the limitations set by fluctuations 
due to extragalactic sources on Planck measurements of primordial CMB 
anisotropies.
On the other hand, the multifrequency all sky surveys carried out by the 
Planck Surveyor mission will provide a very rich database for 
astrophysical studies; their impact on investigations of physical 
and evolutionary properties of different classes of extragalactic sources 
is briefly outlined in \S$\,$3. Our main conclusions are presented in \S$\,$4.

\section{Small scale fluctuations due to extragalactic sources}

A detailed discussion of the problem has been recently carried out by 
Toffolatti et al.~\cite{Toffo}. Guiderdoni et al.~\cite{Guider1} 
have worked out semi-analytic models for galaxy evolution in the IR/sub-mm 
range and have presented predictions for source counts in the Planck/HFI 
bands, among others. The source confusion noise in the same bands 
has also been estimated by Blain et al.~\cite{Blain}.

Due to the steep increase with frequency of the dust emission spectrum 
in the mm/sub-mm region 
(the typical spectral index is $\alpha \simeq -3.5$, $S_\nu 
\propto \nu^{-\alpha}$), 
the crossover between radio and dust emission components for essentially 
all classes of extragalactic sources occurs at wavelengths of a few 
mm; dust temperatures tend to be higher for distant high luminosity 
sources, partially compensating the effect of redshift. 
The minimum in the spectral energy distribution of sources is thus 
roughly coincident with the CMB intensity peak, making the mm region 
ideal for mapping primordial anisotropies. 
A further consequence is that there is an abrupt change in the source 
populations observed in channels above and below 
$\sim 1\,$mm: radio sources dominate at longer wavelengths, while 
in the sub-mm region the Planck instruments will mostly see 
dusty galaxies. 

\subsection{Radio sources}

Estimates of the confusion noise due to radio sources do not 
require extrapolations in flux: 
the available source counts are more than sensitive  enough to include any 
radio source of the familiar steep and ``flat''-spectrum classes, 
likely to cause detectable fluctuations in any of the Planck maps.
The real issue is the spectral behaviour since existing surveys 
extend only up to 8.4 GHz and hence substantial extrapolations  
in frequency are required.  

The results reported by Toffolatti et al.~\cite{Toffo} 
are based on evolutionary models fitting the observed counts, 
as well as redshift and luminosity distributions, of both steep- and 
``flat""-spectrum 
radio sources at several frequencies up to 8.44 GHz. 
In the Planck channels one has to deal primarily with 
``flat''-spectrum sources, mostly at substantial redshifts. 
These sources appear to keep a spectral index $\alpha \sim 0$ 
 up to $\simeq 100\,$GHz \cite{Impey} 
\cite{Edelson}. At higher frequencies a steepening or even a spectral break 
is generally observed \cite{Landau}. 
Considerable uncertainties however remain on the high frequency behaviour 
of these sources. To allow for this, Toffolatti et al.~\cite{Toffo} have 
considered three different values for the mean spectral index 
of compact radio sources in the range 20--200 GHz: $\alpha =-0.3$, 0, and 
0.3; above 200 GHz a steepening to $\alpha=0.7$ was assumed.

\subsection{Evolving dusty galaxies}

We are faced here with the need of substantial extrapolations 
both in frequency and in flux.  
In fact, there is a wide gap with the nearest wavelength ($60\,\mu$m) where 
the most extensive (yet relatively shallow) surveys exist. There is a 
considerable spread in the distribution of dust temperatures, so that 
the $1.3\,\hbox{mm}/60\,\mu$m flux ratios of galaxies span 
about a factor of 10 \cite{Chini} \cite{FranAnd} \cite{DeZotti}.
A tentative estimate of the luminosity function of galaxies at mm wavelengths 
based on a $1.25\,\hbox{mm}/60\,\mu$m bivariate luminosity distribution
has been presented by Franceschini et al. \cite{FRD}. 
Furthermore, the observational constraints on evolution of 
far-IR sources are very poor. IRAS $60\,\mu$m 
counts cover a limited range in flux and are rather uncertain 
at the faint end \cite{Hacking} \cite{Gregor} \cite{Bertin}. 

From a theoretical point of view, there is a great deal of uncertainty on 
the physical processes governing galaxy formation and evolution. Some 
models assume that the comoving density of galaxies remained 
constant after their formation, while they evolved in luminosity due to 
the ageing of stellar populations and the birth of new generations of stars 
(pure luminosity evolution). 
On the other hand, according to the hierarchical galaxy formation
paradigm, big galaxies are formed by coalescence of large numbers
of smaller objects. 
Furthermore, evolution depends on an impressive number of unknown or
poorly known parameters: merging rate, star formation rate, initial 
mass function, galactic winds, infall, interactions, dust properties, 
etc.

Although the evolutionary history is highly uncertain, strong evolution
is expected in the far-IR/mm region particularly 
for early type galaxies since during their early phases they must 
have possessed a substantial metal enriched interstellar medium. 
This expectation is supported by evidences of large amounts of dust
at high redshifts \cite{Omont} \cite{Mazzei} \cite{Hughes}, 
and by the intensity of the isotropic sub--mm 
background \cite{Puget} \cite{Guider} \cite{Schlegel} \cite{Hauser}, 
first discovered by Puget et al.~\cite{Puget}. 
Moreover, the large, negative K-correction strongly amplifies the 
evolutionary effects, that may thus be appreciable in the relatively 
shallow Planck surveys, at least in the highest frequency bands.

Toffolatti et al.~\cite{Toffo} adopted the luminosity evolution models by 
Franceschini et al. \cite{Fran94}, updated adding a density evolution up to 
$z=2$ of late type galaxies \cite{Fran97}. 
Their reference model (referred to as updated model C), 
which entails a dust enshrouded phase during the 
early evolution of spheroidal systems (early type galaxies and bulges of 
disk galaxies), provides a good fit 
to the $60\,\mu$m IRAS counts, as recently reassessed by Bertin et al. 
\cite{Bertin}, to the far-IR extragalactic background spectrum, 
and to the preliminary estimates of the counts 
at $170\,\mu$m \cite{Kawara} and at $850\,\mu$m \cite{Smail}. 

A different approach, directly plugged in current hierarchical 
scenarios for galaxy formation, has been taken by Guiderdoni et al. 
\cite{Guider1} who produced a set of models starting 
from a description of non-dissipative and dissipative collapses of primordial 
perturbations.  
The most extreme of these models exceed, even by 
an order of magnitude, predictions of the updated model C. An extreme 
evolution in the far-IR to sub-mm bands is indeed indicated by the 
most recent estimates \cite{Schlegel} \cite{Hauser} 
indicating a very intense far-IR extragalactic background. 


The contributions of AGNs to the counts in the Planck high frequency
channels are even more uncertain. The detailed estimates by Granato et al. 
\cite{Granato} indicate 
a detection rate of radio quiet AGNs increasing 
with increasing frequency, from a few units over the whole sky at 217 GHz 
to a few hundreds at the highest frequencies; in any case, their number 
is expected to be small compared with the 
number of far-IR galaxies.

\subsection{Fluctuations due to extragalactic sources and CMB anisotropy 
measurements}

\begin{figure}
\psfig{file=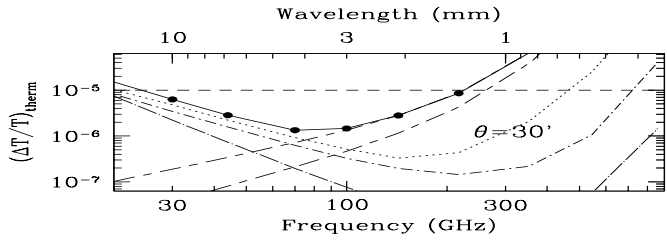,width=12truecm,height=7.5truecm}
\vspace*{-3.5truecm}
\caption{Temperature fluctuations as a function of frequency for an angular 
scale of $30'$. The horizontal dashed line show the expected level 
of primordial CMB anisotropies. Average contributions from Galactic 
free-free, synchrotron and dust (for the two 
possible dust emission spectra mentioned by Kogut et al.{\ }$^{31}$) 
emissions at $|b|> 50^\circ$ are shown by dots/short dashes, 
dots-long dashes, long/short dashes, respectively. The dotted line 
gives the contribution of extragalactic sources. The solid line is the total 
contribution of astrophysical foregrounds to fluctuations; the filled 
circles on this line correspond to Planck channels.
\label{Figdt_30}}
\end{figure}

\begin{figure}
\psfig{file=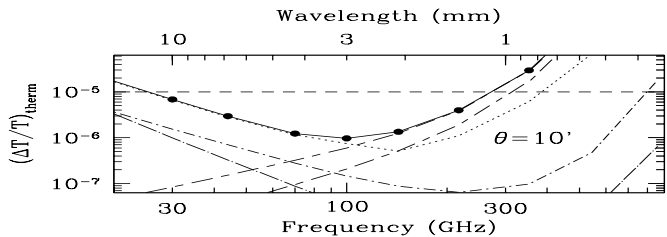,width=12truecm,height=7.5truecm}
\vspace*{-3.5truecm}
\caption{Same as in Fig.~1 but 
for an angular scale of $10'$. 
\label{Figdt_10}}
\end{figure}

Given the performances 
of the instruments, the Planck experimental accuracy is effectively limited 
by astrophysical foregrounds. It is true that, on small 
angular scales ($\theta < 30'$), on average, the instrumental 
noise may exceed fluctuations due to extragalactic sources particularly 
if these are subtracted out down to relatively faint flux levels 
(on larger scales 
the dominant foreground fluctuations are due to Galactic emission and 
generally exceed the instrumental noise). However, much lower 
than average instrumental noise levels, well below the amplitude 
of extragalactic source fluctuations, 
will be reached in regions around the 
ecliptic poles that will be scanned many times; these regions are 
large enough to allow a careful determination of the power spectrum of CMB 
anisotropies on small scales.
On moderate to large angular scales ($\theta \gsim 30'$), foreground 
fluctuations are minimum at frequencies $\simeq 70\,$GHz 
(see Fig.~\ref{Figdt_30}) while on smaller scales the minimum moves 
to $\simeq 100\,$GHz (Fig.~\ref{Figdt_10}).

In the frequency range 100--200 GHz  
CMB anisotropies are expected to dominate over foreground fluctuations 
at least up to $\ell \simeq 1000$.  Only a tiny fraction of 
pixels is expected to be contaminated by extragalactic sources which 
can be efficiently identified thanks to the multifrequency observations 
carried out during the mission.
If only the brigthest ($S > 1\,$Jy) sources are removed, the 
contribution of clustering to fluctuations is small 
compared with the Poisson term. However, the effect of clustering 
may show up in the case of source subtraction down to below 100 mJy, 
introducing a modest feature in the angular power spectrum at 
$\theta \sim 20'$ at high frequencies (clustering of galaxies) and  
$\theta \sim 80'$ at low frequencies (clustering of radio sources).

The separation of CBR fluctuations 
from those due to extragalactic sources is further eased by the 
substantial difference between the power spectra of the two components 
\cite{Tegmark}. 
Also, in principle, it is possible to discriminate fluctuations due 
to discrete sources from primordial CMB anisotropies on the basis 
of their quite different angular size, by means of higher resolution 
observations.
Thus, contamination by extragalactic sources does not set a critical 
limitation to Planck mapping of primordial CMB anisotropies.

\section{Studies of extragalactic sources with the Planck Surveyor mission} 



\subsection{``Flat''-spectrum radio sources}

According to the estimates by Toffolatti et al. \cite{Toffo} the 
Planck surveys will provide 
multifrequency data for a relatively large 
(from several hundred to a few thousand sources), complete sample of 
``flat-''spectrum radio sources (compact radio galaxies, radio loud QSOs, 
BL Lacs, blazars), 
allowing investigation of a number of interesting issues. 

The available 
observations of the spectral energy distributions of these sources 
generally have a gap at mm/sub-mm wavelengths. 
Those sources which have data in this interval  
frequently show a dip in the mm region, indicative of a cross-over of 
two components. 
This spectral feature, that Planck can observe, carries  
a good deal of extremely interesting information on the physical 
properties of sources. 
For example, in flow models of compact radio sources, the spectrum 
steepens at a frequency at which the radiative cooling time 
equals the outflow time \cite{Begel}; for ``hot spots'', 
this typically lies in the millimeter or far-IR part of the 
spectrum, while, in cocoons or extended regions of lower surface brightness, 
the break moves down to lower frequencies.   

According to the basic model \cite{Blandford} \cite{Scheuer}, 
which has been supported by a large body of observational evidence, 
the spectral break frequency, $\nu_b$, at which the synchrotron 
spectrum steepens, is related to the magnetic field $B$ 
and to the ``synchrotron age'' $t_s$ (in Myr) by $\nu_b \simeq 96 
(30\,\mu\hbox{G}/B)^{3}t_s^{-2}\,$GHz. 
 Various evolutionary models of 
the radio emission spectrum have been proposed based on different 
assumptions \cite{Myers}   
(``one-shot'' or continuous injection of relativistic electrons,  
complete or no isotropization of the pitch-angle distribution). 
These models 
strongly differ in the form of the falloff above $\nu_b$.
Also, many compact sources are observed to become optically thin 
at $\nu \gsim 10\,$GHz. Correspondingly, 
their spectral index steepens to values 
($\alpha \simeq 0.7$) typical of extended, optically thin sources. 
Thus, the systematic multifrequency 
study at the Planck frequencies will provide a 
statistical estimate of the radio source ages and measurements of 
the high frequency spectral behaviour and 
of its evolution with cosmic time: these are pieces of information of 
great physical importance. 


In the case of blazars \cite{Brown} the 
component dominating at cm wavelengths is rather ``quiescent'' (variations 
normally occur on time of years) and has a spectral turnover 
at $\sim 2$--5 cm, where the transition between optically thick and 
optically thin synchrotron emission occurs. At higher frequencies 
the emission is dominated by a violently variable ``flaring'' component, 
which rises and decays on timescales of days to weeks, and has a 
self-absorption break at mm/sub-mm wavelengths. The mm/sub-mm region 
is thus crucial to understanding the mechanisms 
responsible for variability in radio loud active nuclei.

It is known from VLBI studies that the apparently smooth ``flat'' spectra 
of compact radio sources are in fact the combination of emissions 
from a number of components with varying synchrotron self absorption 
frequencies which are higher for the denser knots. 
The mm/sub-mm region is unique for studying sub-parsec 
scale, high density regions, including the radio core \cite{Lawrence}.
Thus, while lower frequency surveys provide much more detailed information 
relevant to define {\it phenomenological} evolution properties, surveys 
at mm wavelenghts are unique to provide information on the {\it physical} 
properties.

Excess far-IR/sub-mm emission, possibly due to dust, is often observed from 
local radio galaxies \cite{Knapp}. Planck data will allow to assess 
whether this is a general property of these sources; 
this would have interesting implications 
for the presence of interstellar matter in the host galaxies, 
generally identified with giant ellipticals, which are 
usually thought to be devoid of interstellar matter.

\subsection{Inverted-spectrum radio sources}

The predictions of Toffolatti et al. \cite{Toffo} do not explicitly include 
sources with strongly inverted spectra, peaking at mm wavelengths, 
that would be either missing from, or strongly 
under-represented in low 
frequency surveys and be very difficult to distinguish 
spectrally from fluctuations in the CMB \cite{Crawford}. 

GHz Peaked Spectrum radio sources (GPS) appear to have 
a fairly flat distribution of peak frequencies extending out to 15 GHz in 
the rest frame \cite{ODea}, suggesting the existence of 
an hitherto unknown population of sources with peak at high frequency 
\cite{Lasenby}. 
It is very hard to guess how common such sources may be. Snellen~\cite{Snellen} 
exploited the sample of de Vries et al. \cite{deVries} to estimate a count 
of $22\pm 10\,\hbox{Jy}^{-3/2}\,\hbox{sr}^{-1}$ for sources having 
peak frequencies between 1 and 8 GHz and peak flux densities between 
2 and 6 Jy. He also found that counts of GPS sources are only slowly 
decreasing with increasing peak frequency in that range. If indeed 
the distribution of peak frequencies extends up to several tens GHz 
keeping relatively flat,  
it is conceivable that from several tens to hundreds of GPS sources 
will be detected by the Planck experiment. 
Thus, although these rare sources will not be a threat for studies 
of CMB anisotropies, we may expect that the Planck surveys 
will provide crucial information about their 
properties. GPS sources are important 
because they may be the younger stages of radio source evolution 
\cite{Fanti} \cite{Readhead} and may thus provide insight 
into the genesis and evolution of radio sources; alternatively, they 
may be sources which are kept very compact by unusual conditions 
(high density and/or turbulence) in the interstellar medium of the 
host galaxy \cite{vanBreu}. 

Planck/LFI may also allow to study  
another very interesting class of radio sources, 
powered by advection-dominated accretion flows \cite{NarayanY} \cite{FabRee} 
\cite{DiMatteo} \cite{Fran98}. These may correspond to the final stages 
of accretion in large elliptical galaxies hosting a massive black hole.
Their radio emission is characterized by an inverted spectrum
with spectral index $\alpha \sim -0.4$ up to a frequency of 100--200 GHz,
followed by fast convergence. 

\subsection{Evolving dusty galaxies}

As shown by Toffolatti et al. \cite{Toffo} and 
Guiderdoni et al. \cite{Guider1}, the Planck high frequency 
channels will detect the dust emission from 
a large number of evolving galaxies.
A spectacular breakthrough has been achieved in the last couple 
of years, in the optical/UV,  
with the long sought detection of large samples of galaxies at $z\sim 3$, 
allowing to get a direct insight into the history of the cosmic star and 
metal formation. On the other hand, 
several lines of evidence indicate that optical/UV data are offering a very 
incomplete view of the galaxy evolution at high $z$ \cite{Guider} 
\cite{Burigana} \cite{DeZotti1} \cite{Rowan}. Indeed, there are indications 
that most of the 
starlight emitted during early phases of the evolution of 
spheroidal systems may be essentially invisible in the optical-UV 
region \cite{Guider} \cite{DeZotti}. This may not 
be surprising since active star formation is generally observed to 
occur in dusty environments.
The very strong, negative K-correction due to the steep rise with 
frequency of the dust emission spectrum in the mm/sub-mm wavelength range, 
makes this spectral region particularly well suited for detecting 
high-$z$ galaxies. In fact, models \cite{Guider1} predict that 
most sources detected in the Planck high frequency channels are 
at $z \geq 1$; in some cases, substantial tails up to $z\simeq 3$--4 
are expected. 
Planck high frequency surveys may thus provide a wealth of data essential 
to understand the the cosmic star and metal formation history.

\section{Conclusions}

Luckily enough,  both for galaxies and active galactic nuclei, 
the crossover between the radio and the dust emission 
components, determining a minimum in the spectral energy distribution, 
is roughly coincident with the CMB intensity peak. The dust temperature 
tends to be higher for bright distant objects, moving the minimum 
to higher frequencies in the rest frame and thus partially compensating 
for the effect of redshift. This situation makes the 
mm region ideal for mapping primordial anisotropies. 

Although our understanding of foregrounds at Planck frequencies is far from 
complete, estimates using worst-case parameters in extrapolating 
existing measurements to Planck frequencies or angular scales, allow us 
to safely conclude that,  
in the frequency range 100-200 GHz, the foreground fluctuations, which 
are dominated, on small scales ($\theta \lsim 30'$), by extragalactic sources, 
are well below the expected amplitude of CMB anisotropies over 
much of the sky. Hence, the removal of foreground contamination is not 
critical for accurate determinations of the power spectrum of CMB 
anisotropies up to multipoles of at least $\ell \sim 1000$. 

On the other hand, while only a small fraction of high Galactic latitude 
pixels are strongly contaminated by astrophysical foregrounds,  
the Planck surveys at 9 frequencies will provide sufficiently rich 
complete samples for astrophysical studies. Spectral information will 
be provided for ``flat''-spectrum radio sources (compact radio galaxies, 
radio loud QSOs, BL Lacs, blazars) over a frequency region where 
spectral features carrying essential information on their physical 
conditions show up (breaks due to energy losses of relativistic 
electrons, self-absorption turnovers of flaring components, ...). 
Planck surveys will be unique in providing complete samples of 
bright radio sources with inverted spectra, essentially undetectable 
in radio-frequency surveys. Important classes of sources of this kind 
are GHz peaked spectrum sources, which may be the youngest stages of 
radio source evolution and may thus provide insight into the genesis of 
radio sources, and advection dominated sources, corresponding to 
final stages of accretion in giant elliptical galaxies hosting a 
massive black hole. The high frequency Planck channels will detect 
thousands of dusty galaxies, a large fraction of which at substantial 
redshifts, allowing to extensively test scenarios for galaxy evolution. 
The increasing evidence that a large, and perhaps dominant fraction, 
of star formation at high redshifts may be hidden by dust, makes 
far-IR to sub-mm surveys an essential complement to optical data. 

\section*{Acknowledgments} GDZ acknowledges 
enlightening discussions with R. Fanti on GHz peaked sources and with 
A. Franceschini on advection dominated sources. We are benefited 
from many years of collaboration on foregrounds with L. Danese, C. Burigana, 
A. Franceschini, and P. Mazzei. Work supported in part by CNR and ASI.

\end{document}